\documentclass[english,prl, two column]{revtex4-1}
\usepackage[T1]{fontenc}
\usepackage[latin9]{inputenc}
\usepackage{babel}
\usepackage{float}
\usepackage{amstext}
\usepackage{graphicx}
\usepackage{esint}
\usepackage[unicode=true,pdfusetitle,
 bookmarks=true,bookmarksnumbered=false,bookmarksopen=false,
 breaklinks=false,pdfborder={0 0 1},backref=false,colorlinks=false]
 {hyperref}
\hypersetup{
 colorlinks,linkcolor=red,citecolor=blue}
\usepackage{breakurl}
\begin{document}

\title{Noise-free quantum optical frequency shifting driven by mechanics}

\author{Linran Fan, Chang-Ling Zou, Menno Poot, Risheng Cheng, Xiang Guo,
Xu Han}

\affiliation{Department of Electrical Engineering, Yale University, New Haven,
Connecticut 06511, USA}

\author{Hong X. Tang}

\email{hong.tang@yale.edu}

\affiliation{Department of Electrical Engineering, Yale University, New Haven,
Connecticut 06511, USA}

\maketitle
\textbf{The ability to manipulate single photons is of critical importance
for fundamental quantum optics studies and practical implementations
of quantum communications. While extraordinary progresses have been
made in controlling spatial, temporal, spin and orbit angular momentum
degrees of freedom \cite{key-1,key-3,key-4,key-5,key-6,key-7}, frequency-domain
control of single photons so far relies on nonlinear optical effects,
which have faced obstacles such as noise photons, narrow bandwidth
and demanding optical filtering \cite{key-8,key-9,key-10,key-11,key-12,key-13,key-14,key-15}.
Here we demonstrate the first integrated near-unity efficiency frequency
manipulation of single photons, by stretching and compressing a waveguide
at 8.3 billion cycles per second. Frequency shift up to 150$\,$GHz
at telecom wavelength is realized without measurable added noise and
the preservation of quantum coherence is verified through quantum
interference between twin photons of different colors. This single
photon frequency control approach will be invaluable for increasing
the channel capacity of quantum communications and compensating frequency
mismatch between quantum systems, paving the road toward hybrid quantum
network.}

Photons are essential information carriers for quantum communications
over long distance, benefiting from low dissipation and decoherence
rates at room temperature due to weak interactions with environment.
While unitary operations of single photons on temporal, spatial and
polarization domains can be readily realized with passive photonic
devices \cite{key-1,key-3,key-4,key-5}, frequency control requires
active photonic devices to change photon energy, which are challenging
due to the weak photon-photon interaction \cite{key-8,key-9,key-10,key-11,key-12,key-13,key-14,key-15}.
Nonetheless, frequency control of single photons plays a central role
for a wide range of quantum technologies \cite{key-10,key-11,key-12,key-13,key-14,key-15,key-16}.
It provides an efficient approach to encode quantum information through
frequency multiplexing, thus can boost quantum channel capacity \cite{key-16}.
In addition, the frequency control of single photons is indispensable
to bridge different quantum systems. Frequency mismatch resulting
from disparate qubit systems \cite{key-17}, qubit inhomogeneity \cite{key-19}
and qubit spectrum diffusion \cite{key-20}, can be resolved through
frequency manipulation. 

Typically, single photon frequency conversion is realized with nonlinear
optical processes \cite{key-8,key-9,key-10,key-11,key-12,key-13,key-14,key-15}
and effective nonlinear optical effects mediated by intermediate excitations
\cite{key-21,key-22}. Because of the inherent weak photon-photon
interaction in dielectrics, strong optical pump fields are required
to stimulate the frequency conversion \cite{key-8,key-9,key-10,key-11,key-12,key-13,key-14,key-15}.
Therefore, high-performance optical filtering at single photon levels
is essential to separate signals from strong optical pump fields and
noise photons generated through undesired nonlinear optical effects
such as Raman and fluorescence \cite{key-23}. Moreover, noise photons
in the same spectrum range with signals cannot be filtered. Also,
the frequency conversion based on nonlinear optical processes can
only work at certain wavelengths with limited bandwidth, due to the
stringent momentum and energy conservation conditions \cite{key-9,key-10,key-11,key-12,key-13,key-14,key-15}.

In this Letter, we realize frequency manipulation of single photons
through controlling the mechanical motion of a waveguide. By resonantly
stretching a waveguide at 8.3 GHz, we demonstrate frequency shifting
of single photons with near-unity intrinsic efficiency in a fully
integrated device. High-visibility quantum interference between twin
photons with different colors is observed, indicating the quantum
coherence conservation. The noise during the conversion process is
also found to be below measurable level.

\begin{figure}
\begin{centering}
\includegraphics[width=1\columnwidth]{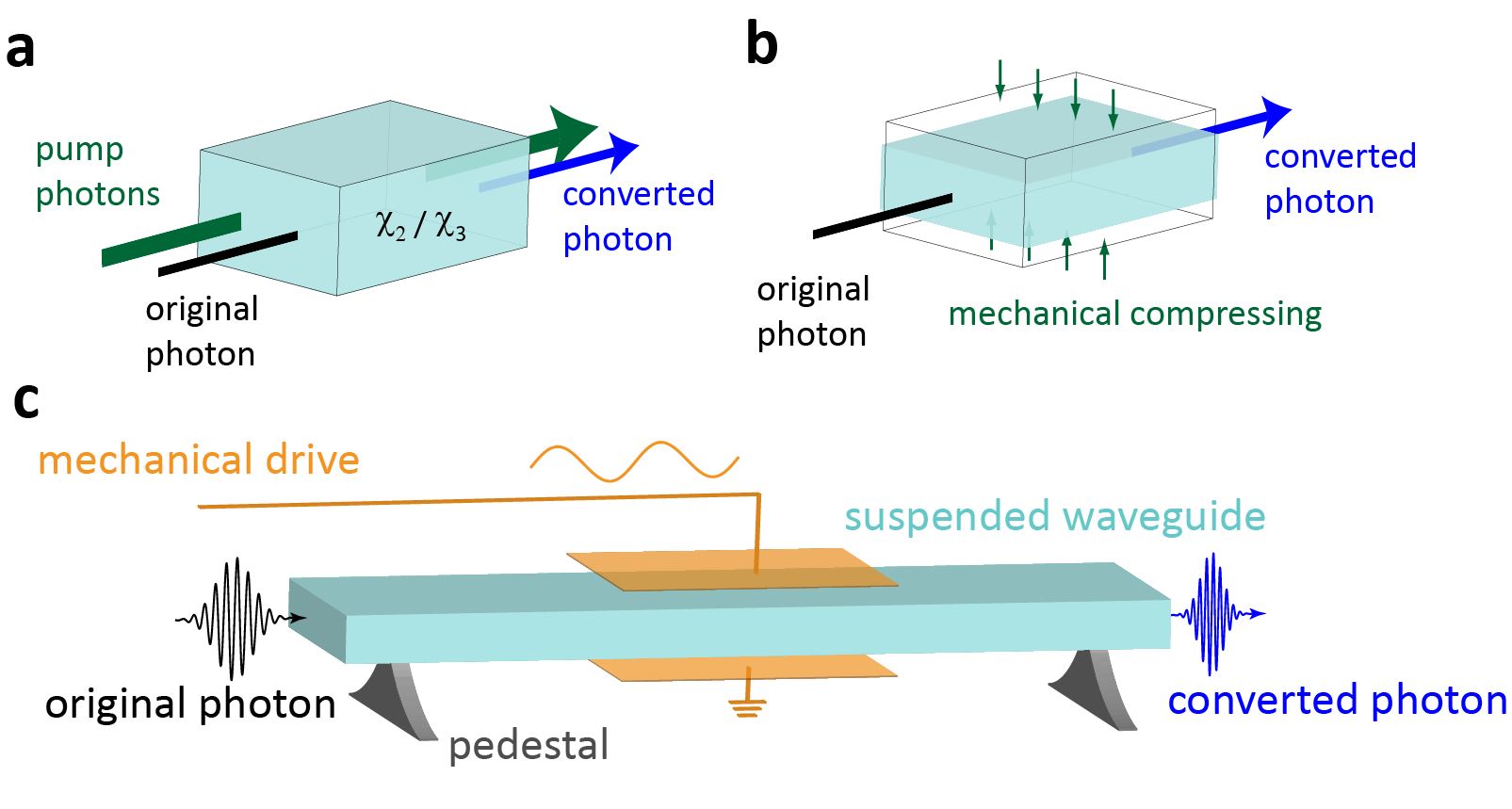}
\par\end{centering}

\caption{\textbf{Principles of quantum frequency conversion. a. }Frequency
conversion using nonlinear optical processes. \textbf{b. }Frequency
conversion induced by the mechanical deformation. During the photon's
propagation, the medium is compressed or stretched rapidly, causing
frequency shift of the photon. \textbf{c. }Schematic of a suspended
waveguide made of piezoelectric material. The co-localization of the
mechanical and optical modes in the waveguide leads to strong optomechanical
interaction, and the piezoelectric effect provides the efficient interface
to control the mechanical motion using RF signals.}

\label{Fig1}
\end{figure}

The principle of conventional frequency conversion based on nonlinear
optical processes is illustrated in Fig.$\,$\ref{Fig1}a, where input
photons are mixed with pump photons in nonlinear optical material
to generate output photons at another frequency. Figure$\,$\ref{Fig1}b
explains our approach using mechanical deformation. The mechanical
deformation changes optical length, leading to compression or stretching
of photons, which results in frequency shift. In contrast to multi-wave
mixing in Fig.$\,$\ref{Fig1}a, the frequency conversion induced
by the mechanical deformation does not require optical pumps, thus
noise generation mechanisms arising from high intensity optical pump
are eliminated. This process can also be realized with electro-optic
effect, and here the frequency shift induced by mechanics can be further
enhanced by the mechanical resonance. In the adiabatic limit where
the mechanical change rate is much smaller than the optical frequency
$\omega_{\mathrm{o}}$, the frequency shift $\delta\omega$ is proportional
to the mechanical deformation $\delta x$ 
\begin{equation}
\delta\omega\approx-v_{\mathrm{g}}k_{\mathrm{o}}\frac{\partial n_{\mathrm{eff}}}{\partial x}\delta x,\label{eq:1}
\end{equation}
where $k_{\mathrm{o}}$, $v_{\mathrm{g}}$ and $n_{\mathrm{eff}}$
are the photon vacuum wave-vector, the group velocity and the effective
refractive index respectively (Supplementary Section I). In such an
adiabatic process, the adiabatic invariant is the photon number, thus
photons can be converted with unity efficiency \cite{key-24,key-25}.
Adiabatic frequency shifting has also been demonstrated in optical
micro-cavities based on refractive index modulation through carrier
injection \cite{key-25,key-26,key-27}. Although impressive progresses
have been achieved, the cavity-based structure and carrier-injection
mechanism are still in need of significant improvement to overcome
their limits in efficiency, noise and bandwidth.

\begin{figure}
\begin{centering}
\includegraphics[width=1\columnwidth]{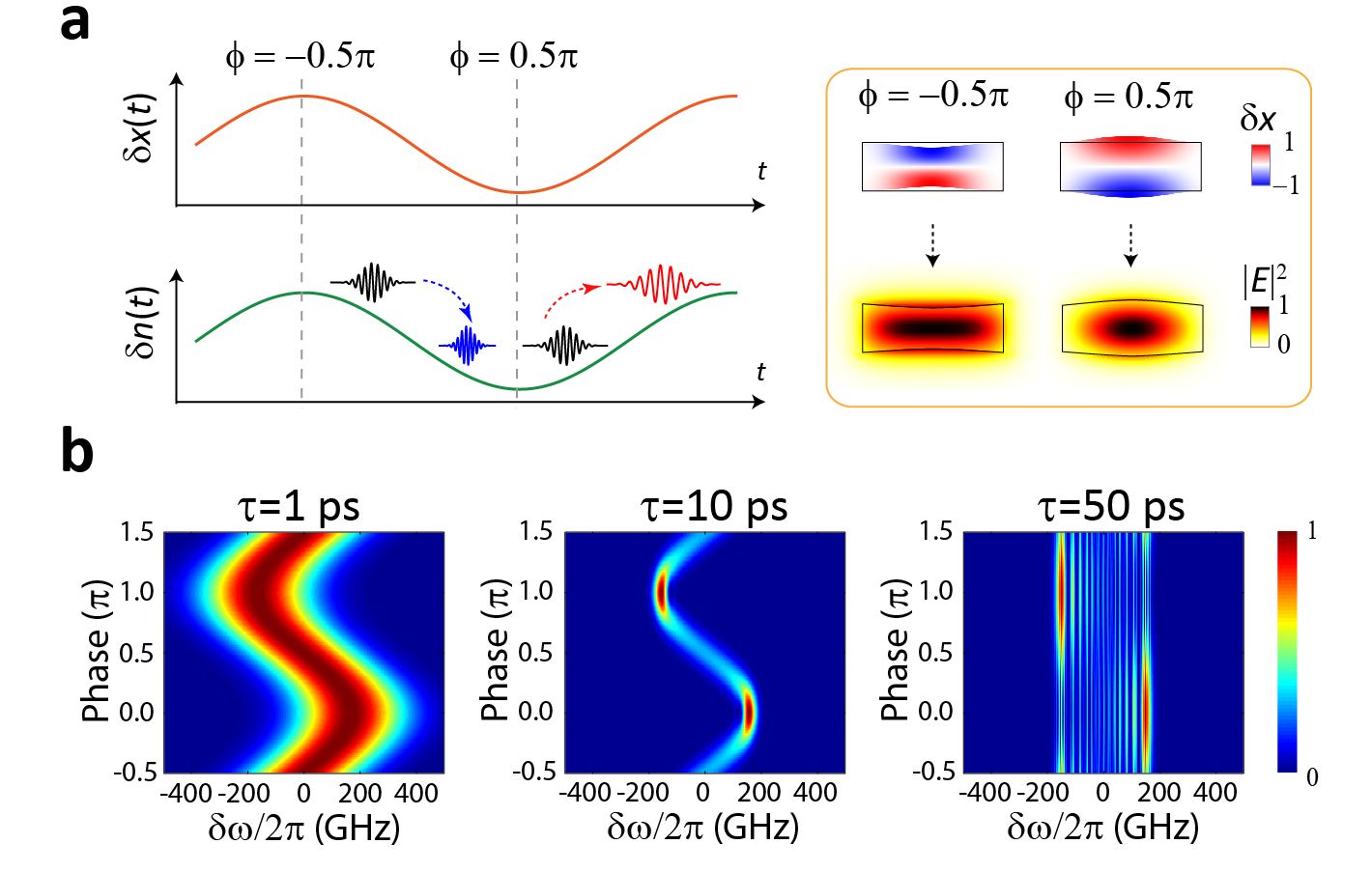}
\par\end{centering}

\caption{\textbf{Frequency conversion induced by the mechanical motion.} \textbf{a.}
With different mechanical oscillation phases $\phi$, photons experience
different refractive index changes, and thus different frequency shift
directions and amplitudes. The mechanical thickness mode and TM optical
mode of a rectangular waveguide cross-section are shown. The positive
(negative) displacement leads to the increase (decrease) of effective
refractive index change. \textbf{b.} Simulated output spectrum of
a Gaussian input photon (i.e. $a(t)=e^{-i\omega_{\mathrm{\mathrm{o}}}(t-t_{0})^{2}/\tau^{2}}$)
(Supplementary section I). The parameters used are the waveguide length
$L=5\,\textrm{mm}$, the refractive index modulation depth $\delta n_{\mathrm{eff}}/n=5\times10^{-4}$,
and the mechanical frequency $\omega_{\mathrm{m}}/2\pi=8.3\,\textrm{GHz}$,
the group index of TM optical mode $n=2$ and the optical frequency
$\omega_{\mathrm{\mathrm{o}}}/2\pi=194\,\textrm{THz}$. Different
photon durations $\tau$ of 1$\,$ps, 10$\,$ps and 50$\,$ps are
shown.}

\label{Fig2}
\end{figure}

To implement the mechanical adiabatic frequency shifting, we use a
suspended waveguide made of piezoelectric material, which provides
efficient, high amplitude actuation of mechanical motion through RF
signals (Fig.$\,$\ref{Fig1}c) \cite{key-28}. The mechanical thickness
mode of the suspended waveguide is used for its high oscillation frequency
and strong interaction with the transverse-magnetic (TM) optical mode
(Fig.$\,$\ref{Fig2}a). Following the mechanical vibration, the effective
refractive index changes periodically. Thus photons in the waveguide
experience different optical length changes with different drive phase
$\phi$ with respect to the photon arrival time. The corresponding
frequency shift can be expressed as
\begin{equation}
\delta\omega=-v_{\mathrm{g}}k_{\mathrm{o}}\frac{\partial n_{\mathrm{eff}}}{\partial x}\delta X[\sin(\phi-\frac{\omega_{\mathrm{m}}L}{2v_{\mathrm{g}}})-\sin(\phi+\frac{\omega_{\mathrm{m}}L}{2v_{\mathrm{g}}})],\label{eq:2}
\end{equation}
where $\omega_{\mathrm{m}}$ is the angular frequency of the mechanical
mode, $\delta X$ is the mechanical motion amplitude, and $L$ is
the waveguide length (Supplementary Section I). In order to avoid
spectral distortion, the photon duration $\tau$ should be much smaller
than the mechanical oscillation period $2\pi/\omega_{\mathrm{m}}$.
In Fig.$\,$\ref{Fig2}b, we compare the simulated output spectrum
of a Gaussian input photon with different $\tau$ and $\phi$. As
predicted by equation \ref{eq:2}, $\delta\omega$ follows a sinusoidal
dependence on $\phi$ for short photon duration. The maximum frequency
shift and minimum shape distortion happen at $\phi=0$ and $\phi=\pi$,
where the refractive index change is largest and most uniform. With
a very short photon duration, the photon spectrum remains Gaussian.
When the photon duration becomes comparable to the mechanical oscillation
period, sidebands are generated and the photon spectrum is distorted
due to the nonlinear refractive index change.

\begin{figure*}
\begin{centering}
\includegraphics[width=0.7\paperwidth]{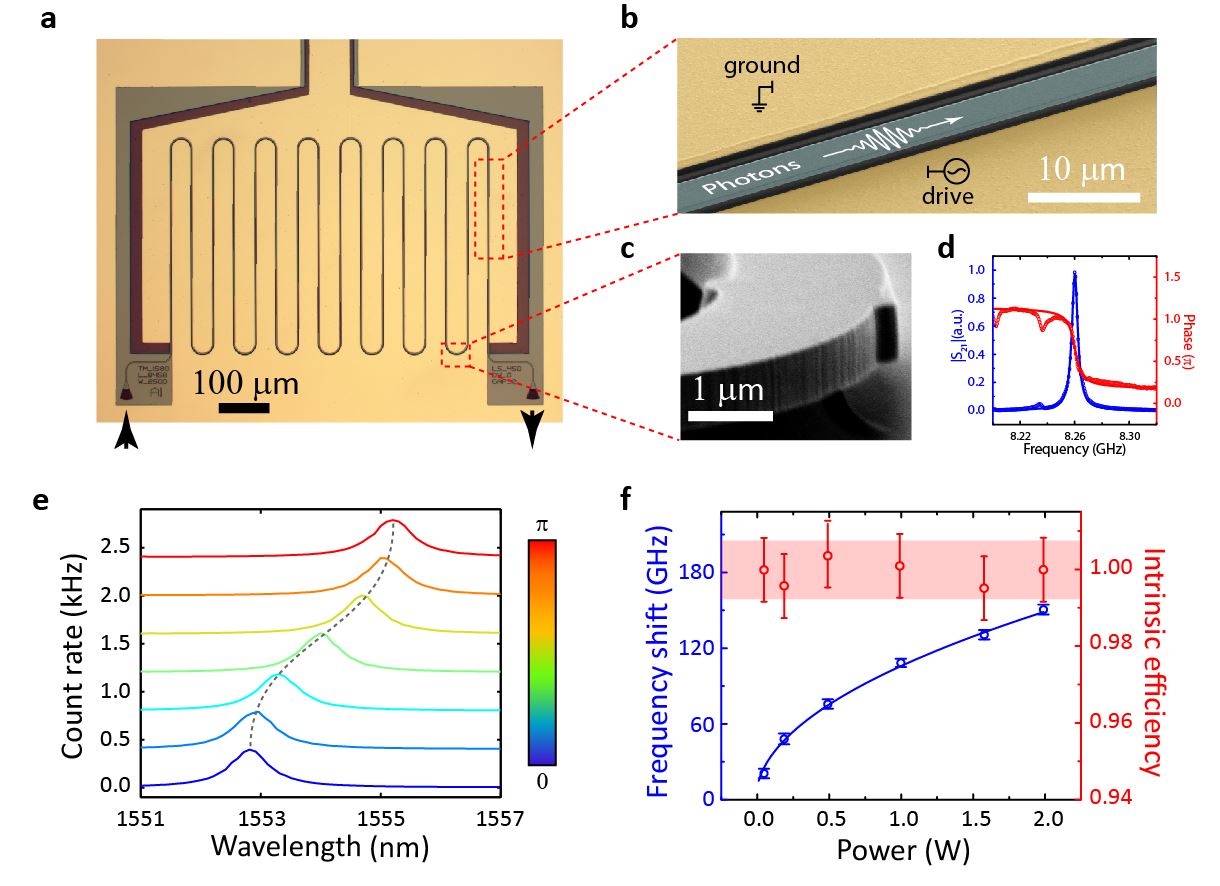}
\par\end{centering}

\caption{\textbf{Single photon frequency control.} \textbf{a.} Optical image
of the suspended AlN waveguide. The waveguide is 6$\,$mm in length,
folded into a meander within $1\times1\,\textrm{m\ensuremath{\textrm{m}^{2}}}$.
\textbf{b.} Scanning electron microscope (SEM) image of the device
in false color. The waveguide (cyan) is 2.4$\,$$\textrm{\ensuremath{\mu}m}$
wide. The horizontal distance between the two RF electrodes (yellow)
and the waveguide are 100$\,$nm and 800$\,$nm respectively. The
two RF electrodes are deposited on top of the AlN and Si layers respectively.
\textbf{c.} SEM image at a supporting point of the waveguide. The
waveguide width is increased by 200$\,$nm, and a pedestal with $100\times100\,\textrm{n\ensuremath{\textrm{m}^{2}}}$size
can be formed to support the waveguide. \textbf{d.} Driven response
of the mechanical thickness mode with a fit to extract the mechanical
quality factor. \textbf{e.} Single photon spectrum measured using
SSPD with an optical bandpass filter. The dotted grey line shows the
theoretical centre wavelength shift (equation \ref{eq:2}). The curves
are offset for clarity.\textbf{ f.} Frequency shift amplitude (blue)
and intrinsic conversion efficiency (red) dependence on the RF drive
power. The solid blue line is the fitted frequency shift with square-root
dependence on the drive power. The shaded area indicates the unity
intrinsic conversion efficiency, measured with reference to the photon
counting rate without the RF drive. The error of the frequency shift
is estimated based on the fitting of the measured spectrum. The error
of the intrinsic conversion efficiency is estimated based on the Possion
distribution of photon counting.}
\label{Fig3}
\end{figure*}

In experiment, the frequency shifting device is fabricated from an
aluminum nitride (AlN) layer on a silicon dioxide (SiO$_{2}$) cladding
on silicon (Si) wafer (Fig.$\,$\ref{Fig3}a and further explained
in Methods). The AlN thickness is 650 nm, resulting in a fundamental
thickness mode frequency of $\omega_{\mathrm{m}}/2\pi=8.26\,\mathrm{GHz}$
(Fig.$\,$\ref{Fig3}d). The waveguide is 6$\,$mm long, folded into
a meander with a footprint of 1$\,$mm$^{2}$. The RF ground and signal
electrodes are situated close to the waveguide, on the AlN and Si
layers respectively to produce strong driving electrical fields (Fig.$\,$\ref{Fig3}b).
The waveguide width is intentionally widened slightly every $100\,\mathrm{\mu m}$
as shown in Fig.$\,$\ref{Fig3}c, where a point-like pedestal can
be formed \cite{key-28}. The pedestal is reduced to around $100\times100\,\textrm{n\ensuremath{\textrm{m}^{2}}}$
in size to minimize the clamping loss, while remaining robust to support
the suspended waveguide. In this way, a mechanical quality factor
over 2000 is achieved at room temperature (Fig.$\,$\ref{Fig3}d).

The device is first tested with photons produced through spontaneous
parametric down conversion (SPDC) with a centre wavelength $1554.0\,\mathrm{nm}$,
and the output photon spectrum is measured by a superconducting single-photon
detector (SSPD) with a $0.3\,\mathrm{nm}$ optical bandpass filter.
The RF drive is synchronized with the photon arrival time to obtain
steady RF phase $\phi$ (Supplementary Section II). The photon spectrum
is shifted without spectral distortion (Fig.$\,$\ref{Fig3}e). By
changing $\phi$ from $0$ to $\pi$ with fixed drive power 2 W, the
centre wavelength is shifted from 1552.8$\,$nm to 1555.2$\,$nm,
corresponding to a maximum frequency shift $\pm2\pi\times150\,\mathrm{GHz}$.
The frequency shift has a sinusoidal dependence on $\phi$, which
agrees with the prediction of equation \ref{eq:2}. Beside the drive
phase, $\delta\omega$ can be precisely controlled by the drive power
as well (Fig.$\,$\ref{Fig3}f). Since the frequency shift is proportional
to the amplitude of the mechanical displacement (equation$\,$\ref{eq:2}),
which in turn is proportional to the square root of the drive power
$P$, we have the frequency shift coefficient $\eta=\frac{\delta\omega}{\sqrt{P}}\approx2\pi\times106.8\,\mathrm{\textrm{GHz/\ensuremath{\sqrt{\mathrm{W}}}}}$.
The intrinsic conversion efficiency can be estimated by comparing
the total count rates under different RF drive power with the count
rate under no RF drive (Fig.$\,$\ref{Fig3}f). No statistically significant
change of count rates is observed, indicating the near-unity intrinsic
conversion efficiency for single photons.

\begin{figure*}
\begin{centering}
\includegraphics[width=0.7\paperwidth]{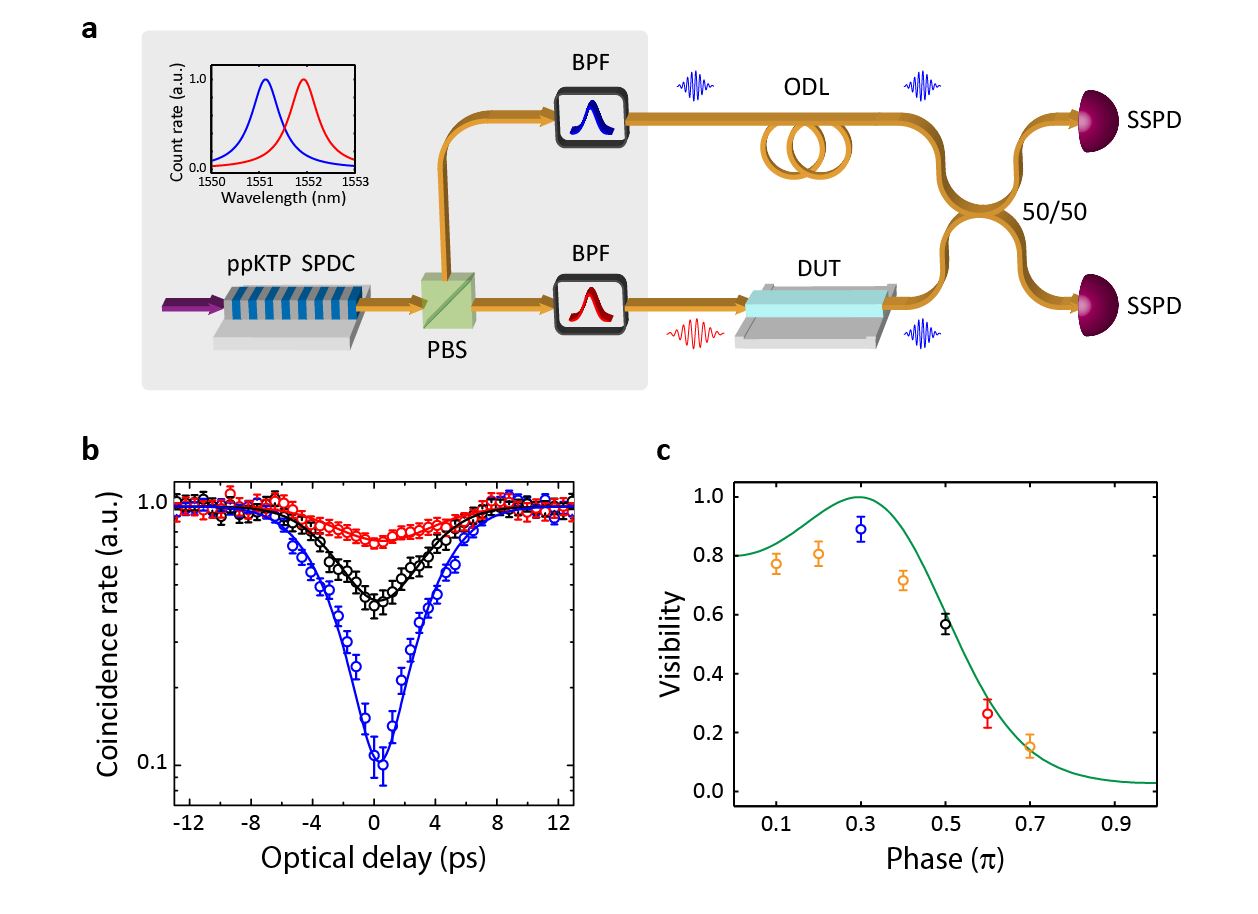}
\par\end{centering}

\caption{\textbf{Two photon quantum interference with different colors. a.}
Experimental setup for two-photon quantum interference with different
colors. ppKTP, periodically-poled potassium titanyl phosphate; PBS,
polarizing beam splitter; BPF, optical bandpass filter; ODL, optical
delay line; DUT, device under test; SSPD, superconducting single-photon
detector. \textbf{b.} Coincidence rate between the two SSPDs normalized
by the coincidence rate at an optical delay much larger than the photon
duration, for unshifted photons (black), red-shifted photons (red)
and blue-shifted photons (blue). \textbf{c. }Visibility dependence
(symbols) on the RF drive phase. The black, red and blue symbols correspond
to the black, red and blue curves in \textbf{b }respectively. Theoretical
visibility (solid line) for different phases with a frequency shift
amplitude of 150$\,$GHz, other parameters are the initial frequency
separation (100$\,$GHz) and the photon duration ($\tau=4\,\mathrm{ps}$).
The error is estimated based on the possion distribution of photon
counting.\textbf{ }}

\label{Fig4}
\end{figure*}

In order to examine the quantum coherence of the conversion process,
we carried out the quantum interference experiment with non-degenerate
photon pairs (Fig.$\,$\ref{Fig4}a) \cite{key-29}. Twin photons
with neglectable spectral correlations respectively centered at 1551.1$\,$nm
(referred to as blue photon) and 1551.9$\,$nm (referred to as red
photon) are generated through type-II SPDC (Supplementary Section
III). The blue and red photons pass through an optical delay line
and the device respectively. Then the photons are coupled into a 50/50
beam splitter, and detected by two SSPDs. The coincidence rate is
calculated by comparing the arrival time lists of the two SSPDs, and
normalized with the coincidence rate at the optical delay much larger
than the photon coherence time (Fig.$\,$\ref{Fig4}b). Without applying
the RF drive, we observe a dip in the coincidence with a visibility
$V=(56.1\pm3.1)\%$ by fitting the coincidence rate with the function
\begin{equation}
C=1-Ve^{-\frac{t^{2}}{2\tau^{2}}},
\end{equation}
where $C$ is the normalized coincidence rate, $V$ is the visibility,
$t$ is the delay time, and $\tau$ is the pulse duration of single
photons \cite{key-30}. By blue-shifting the red photons, the highest
visibility $V=(90.3\pm4.6)\%$ can be reached when the spectra of
the two photons overlap. In contrast, the visibility is reduced by
red-shifting the red photons, indicating the further separation of
the photons in frequency domain. As shown in Fig.$\,$\ref{Fig3}e,
the centre wavelength of photons can be controlled by tuning the drive
phase $\phi$. Therefore, we can change the indistinguishability between
the two photons continuously via the phase $\phi$ (Fig.$\,$\ref{Fig4}c).

The high visibility of the quantum interference confirms the preservation
of quantum coherence. In order to further verify the noise performance,
we also measure the dark count rate of SSPD and the total noise count
rate without input photons and optical filters, which are $42.54\pm0.38\,\mathrm{Hz}$
and $42.39\pm0.38\,\mathrm{Hz}$ respectively. No statistically significant
amount of noise photons from the frequency shifting are measured.
Moreover, the frequency shifting requires no phase matching conditions,
therefore suitable for broadband quantum communication (Supplementary
section II). Compared with bulk electro-optic materials such as LiNbO$_{3}$,
the small footprint, scalability, low insertion loss and CMOS compatibility
make our device easy to integrate into quantum networks. (Supplementary
section IV). Further improvements in frequency shift range can be
readily achieved through integrating more devices on a single chip,
as well as increasing mechanical quality factors by reducing the operation
temperature.

In conclusion, we have demonstrated the frequency manipulation of
single photons by controlling the mechanical motion of integrated
AlN waveguides. With strong optomechanical interaction and piezoelectric
actuation, we have been able to shift single photon frequency over
300$\,$GHz at telecom wavelength with near-unity intrinsic efficiency
and no added noise, while preserving the quantum coherence of single
photons. As no optical pump or phase matching is required in this
process, our results provide an efficient and clean solution for the
single photon frequency control. This paves the way to realize the
high-capacity quantum communications between different quantum systems,
pushing forward the frontier of quantum technology.

\emph{Note added}: Upon submission of this manuscript, we became aware
of the work by Wright \emph{et al.}, where single photon frequency
shifting is realized based on electro-optic effect \cite{key-31}.

\vbox{}

\noindent\textbf{\large{}Methods}{\large \par}

\noindent\textbf{ Device fabrication.} The device is fabricated from
650-nm-thick AlN on $\mathrm{SiO_{2}}$ (thickness $2.2\,\mathrm{\mu m}$)
on a Si wafer. Optical waveguides are patterned with electron beam
lithography (EBL) using hydrogen silsesquioxane (HSQ) resist, subsequently
transferred to the AlN layer by chlorine-based reactive ion-etching
(RIE). In this step, only $500\,\mathrm{nm}$ out of the $650\,\mathrm{nm}$
AlN is initially etched. In a second EBL using ZEP520A resist, the
release window is defined, and followed by RIE of the remaining $150\,\mathrm{nm}$
AlN. A fluorine-based deep RIE is used to selectively etch the $\mathrm{SiO_{2}}$
exposed by the release window down to the silicon substrate. A third
EBL is performed to define the RF electrodes in polymethyl methacrylate
(PMMA). After development, $300\,\mathrm{nm}$ gold is thermally evaporated,
followed by lift-off in acetone. Finally, the waveguides are released
in buffered oxide etchant. By timing the releasing process, the whole
waveguide is suspended, only supported by specifically engineered
pedestals (Fig.$\,$\ref{Fig3}c).

\vbox{}

\noindent\textbf{Acknowledgments}\\ We acknowledge funding support
from an LPS/ARO grant (W911NF-14-1-0563), AFOSR MURI grant (FA9550-15-1-0029),
and DARPA ORCHID program through a grant from the Air Force Office
of Scientific Research (AFOSR FA9550-10-1-0297), and the Packard Foundation.
Facilities used were supported by Yale Institute for Nanoscience and
Quantum Engineering and NSF MRSEC DMR 1119826. The authors thank Michael
Power and Dr. Michael Rooks for assistance in device fabrication.

\vbox{}

\noindent\textbf{Author contributions}\\ H.X.T., L.F. \& C.-L.Z.
conceived the experiment; L.F., R.C. \& X.H. fabricated the device;
L.F., M.P., R.C. \& X.G. performed the measurements; L.F. \& C.-L.Z.
analyzed the data. L.F. \& C.-L.Z. wrote the manuscript, and all authors
contribute to the manuscript. H.X.T. supervised the work.

\vbox{}

\noindent\textbf{Additional information}\\ Supplementary information
is available in the online version of the paper. Reprints and permissions
information is available online at www.nature.com/reprints. Correspondence
and requests for materials should be addressed to H.X.T.

\vbox{}

\noindent\textbf{Competing financial interests}\\The authors declare
no competing financial interests.

\clearpage{}

\newpage{}

\newpage{}

\onecolumngrid
\renewcommand{\thefigure}{S\arabic{figure}}
\setcounter{figure}{0} 
\renewcommand{\thepage}{S\arabic{page}}
\setcounter{page}{0} 
\renewcommand{\theequation}{S.\arabic{equation}}
\setcounter{equation}{0} 
\setcounter{section}{0}

\begin{center}
\textbf{\textsc{\LARGE{}Supplementary Information}}
\par\end{center}{\LARGE \par}

\section{Theory of frequency shifting}

For an optical pulse $a(z,t)e^{i\beta z-i\omega_{0}t}$ propagating
long $z$-direction with centre frequency $\omega_{\mathrm{o}}$ and
wave-vector $\beta$, the pulse propagation equation can be written
as \cite{key-S1}
\begin{equation}
\frac{\partial a(z,t)}{\partial z}+\frac{1}{v_{\mathrm{g}}}\frac{\partial a(z,t)}{\partial t}=i\delta\beta(z,t)a(z,t),\label{eq:S1}
\end{equation}
where $v_{\mathrm{g}}$ is the group velocity, $\delta\beta$ is the
wave-vector change due to external perturbation. For the effective
refractive index change due to harmonically driven mechanical deformation,
we have 
\begin{equation}
\delta\beta(z,t)=\delta n_{\mathrm{eff}}k_{\mathrm{o}}\sin(\omega_{\mathrm{m}}t+\varphi)\Pi(z/L),\label{eq:S2}
\end{equation}
where $\delta n_{\mathrm{eff}}$ is the change of effective refractive
index, $k_{\mathrm{o}}$ is the vacuum wavevector, $\omega_{\mathrm{m}}$
is the driving angular frequency, and $\varphi$ is the mechanical
oscillation phase, $L$ is the waveguide length, and $\Pi(x)$ is
the rectangular function, i.e. $\Pi(x)=0$ for $|x|>\frac{1}{2}$
and $\Pi(x)=1$ for $|x|<\frac{1}{2}$. If we define two new arguments
$\xi=t+\frac{z}{v_{\mathrm{g}}}$ and $\eta=t-\frac{z}{v_{\mathrm{g}}}$,
then equation \ref{eq:S1} becomes
\begin{equation}
\frac{\partial a(\xi,\eta)}{\partial\xi}=i\frac{v_{\mathrm{g}}}{2}\delta\beta(\xi,\eta)a(\xi,\eta),
\end{equation}
which is directly integrable
\begin{equation}
a(\xi,\eta)=g(\eta)\exp\left[i\frac{v_{\mathrm{g}}}{2}\int_{\eta}^{\xi}\delta\beta(x,\eta)dx\right],
\end{equation}
where $g(\eta)$ is determined by the initial photon waveform. Using
equation \ref{eq:S2}, we have 
\begin{equation}
\int_{\eta}^{\xi}\delta\beta(x,\eta)dx=\int_{\eta}^{\xi}\delta n_{\mathrm{eff}}k_{\mathrm{o}}\sin(\omega_{\mathrm{m}}\frac{x+\eta}{2}+\varphi)\Pi(\frac{x-\eta}{2\tau_{\mathrm{p}}})dx
\end{equation}
where $\tau_{\mathrm{p}}=L/v_{\mathrm{g}}$ denotes the pulse propagation
time in the waveguide. For $z=L/2$ at the output port,we have 
\begin{equation}
a(\frac{L}{2},t)=g(t-\frac{\tau_{\mathrm{p}}}{2})\exp\left\{ -iv_{\mathrm{g}}\frac{\delta n_{\mathrm{eff}}k_{\mathrm{o}}}{\omega_{\mathrm{m}}}\left[\cos(\omega_{\mathrm{m}}t+\varphi)-\cos\left[\omega_{\mathrm{m}}(t-\frac{\tau_{\mathrm{p}}}{2})+\varphi\right]\right]\right\} .
\end{equation}
And for $z=-L/2$ at the input port, we have 
\begin{equation}
a(-\frac{L}{2},t)=g(t+\frac{\tau_{\mathrm{p}}}{2})\exp\left\{ -iv_{\mathrm{g}}\frac{\delta n_{\mathrm{eff}}k_{\mathrm{o}}}{\omega_{\mathrm{m}}}\left[\cos\left(\omega_{\mathrm{m}}t+\varphi\right)-\cos\left[\omega_{\mathrm{m}}(t+\frac{\tau_{\mathrm{p}}}{2})+\varphi\right]\right]\right\} .
\end{equation}
Therefore, we obtain the relation between output and input pulses
\begin{equation}
a(\frac{L}{2},t)=a(-\frac{L}{2},t-\tau_{\mathrm{p}})\exp\left\{ -iv_{\mathrm{g}}\frac{\delta n_{\mathrm{eff}}k_{\mathrm{o}}}{\omega_{\mathrm{m}}}\left[\cos(\omega_{\mathrm{m}}t+\varphi)-\cos\left(\omega_{\mathrm{m}}(t-\tau_{\mathrm{p}})+\varphi\right)\right]\right\} .
\end{equation}
If the pulse duration is much smaller than the period of the mechanical
oscillation, i.e. $|t-t_{\mathrm{e}}|\ll\tau_{\mathrm{p}}$ with $t_{\mathrm{e}}$
the time when photon exits the waveguide, the frequency shift can
be approximated as
\begin{equation}
a(\frac{L}{2},t)\approx a(-\frac{L}{2},t-\tau_{\mathrm{p}})\exp(-i\delta\omega t+\psi)
\end{equation}
 
\begin{equation}
\delta\omega=-v_{\mathrm{g}}k_{0}\delta n_{\mathrm{eff}}[\sin(\phi-\frac{\omega_{\mathrm{m}}L}{2v_{\mathrm{g}}})-\sin(\phi+\frac{\omega_{\mathrm{m}}L}{2v_{\mathrm{g}}})]
\end{equation}
where $\phi=\omega_{\mathrm{m}}t_{\mathrm{e}}+\varphi-\frac{\omega_{\mathrm{m}}\tau_{\mathrm{p}}}{2}$
and $\Phi=v_{\mathrm{g}}\frac{\delta n_{\mathrm{eff}}k_{\mathrm{o}}}{\omega_{\mathrm{m}}}\left[\cos(\omega_{\mathrm{m}}t_{\mathrm{e}}+\phi+\frac{\omega_{\mathrm{m}}L}{2v_{\mathrm{g}}})-\cos\left(\omega_{\mathrm{m}}t_{\mathrm{e}}+\phi-\frac{\omega_{\mathrm{m}}L}{2v_{\mathrm{g}}}\right)\right]$.
. 

\begin{figure}[H]
\begin{centering}
\includegraphics[width=0.5\paperwidth]{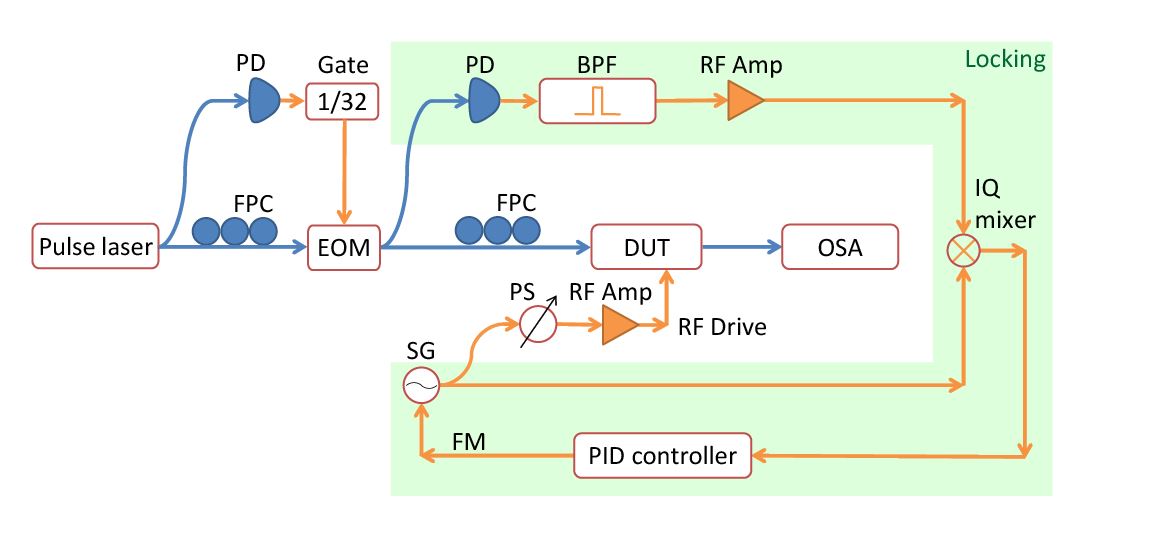}
\par\end{centering}

\caption{\textbf{The experimental setup to realize frequency shifting for optical
pulses. }FPC, fiber polarization controller; PD, photodetector; EOM,
electro-optical modulator; Gate, square wave gate voltage; DUT, device
under test; OSA, optical spectrum analyzer; BPF, bandpass filter;
RF Amp, RF amplifier; FM, frequency modulation; SG, signal generator;
PS, phase shifter.}

\label{FigS2-1}
\end{figure}

\begin{figure}[H]
\begin{centering}
\includegraphics[width=0.5\paperwidth]{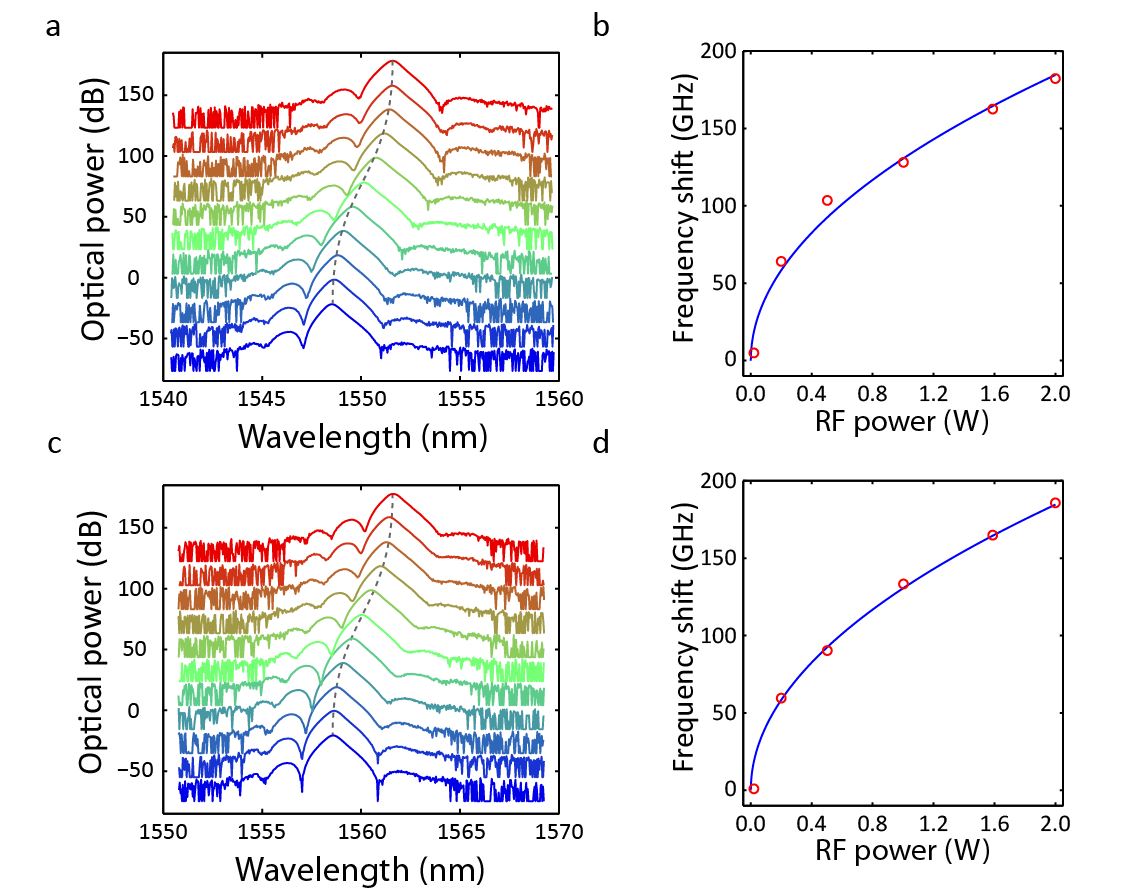}
\par\end{centering}

\caption{\textbf{Frequency shift for optical pulses}. The optical spectrum
after frequency shifting with 1550 nm (\textbf{a}) and 1560 nm (\textbf{c})
input photons is displayed. The phase shifter is tuned from 0 to $\pi$
(blue to red). The dotted grey line shows the theoretical centre wavelength
shift. The frequency shift for both 1550 nm (\textbf{b}) and 1560
nm (\textbf{d}) photons is proportional to the square root of the
RF drive power.}

\label{FigS2}
\end{figure}

\section{Device characterization in the classical regime}

The experimental setup to realize frequency shifting of optical pulses
is displayed in Fig.$\,$\ref{FigS2-1}. We used a mode-lock laser
with 48.6$\,$MHz repetition rate and 4$\,$ps photon duration. First,
the repetition rate is decreased to 1.52$\,$MHz by applying gate
voltage to an electro-optical modulator (EOM). The output from the
EOM is divided into two parts. The first part is sent into the device,
and then the optical spectrum analyzer to monitor the frequency shift.
The second part is sent directly into a high frequency photo-detector.
The optical pulses are short enough to generate high frequency harmonics
of the fundamental 1.52$\,$MHz signal from the photo-detector. One
high-order harmonic signal from the photo-detector is mixed with the
RF drive for the mechanical thickness mode. The mixed signal is used
as the feedback for the RF drive through a PID controller. In this
way, the RF drive is synchronized to the optical pulses from the mode-lock
laser, and the jitter between the arrival time of optical pulses and
RF drive can be controlled below 2$\,$ps, which is much smaller than
the mechanical mode period (\textasciitilde{}120$\,$ps). By synchronizing
the RF drive to different order harmonics of optical pulse, the RF
drive can be stepped by 1.52$\,$MHz, which is fine enough to match
the mechanical resonance (FWHM=4.13$\,$MHz). The synchronized RF
drive is then sent into the device with adjustable phase and maximum
power 33$\,$dBm. 

The centre wavelength of the input optical pulses is set to be 1550$\,$nm.
The measured optical power spectrum after the frequency shifting with
different phase between optical pulses and RF drive is displayed in
Fig.$\,$\ref{FigS2}a. The maximum frequency shift of 180$\,$GHz
is observed on both the red and blue sides. As the mechanical mode
displacement is proportional to the driving electric field strength,
the frequency shift amplitude is proportional to the square root of
the RF drive power as shown in Fig.$\,$\ref{FigS2}b. The fitting
of data gives the shifting coefficient $\eta=\frac{\delta f}{\sqrt{P}}\approx130.8\:\textrm{GHz/\ensuremath{\sqrt{W}}}$,
where $P$ is the RF drive power. Then the centre wavelength of the
input optical pulses is set to be 1560$\,$nm. Similar performance
is observed with the 1560$\,$nm input optical pulses (Fig.$\,$\ref{FigS2}c
\& d).

\section{Experimental setup for two-photon quantum interference with different
colors}

\begin{figure}[H]
\begin{centering}
\includegraphics[width=0.6\paperwidth]{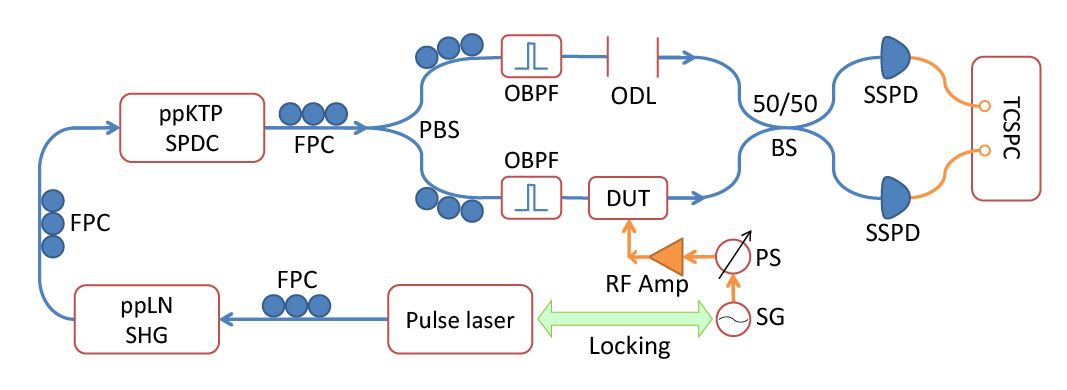}
\par\end{centering}

\caption{\textbf{Experimental setup for two-photon quantum interference with
different colors}. ppLN, periodically poled lithium niobate; ppKTP,
periodically poled potassium titanyl phosphate; FPC, fiber polarization
controller; PBS, polarization beam splitter; OBPF, optical bandpass
filter; ODL, optical delay line; BS, beam splitter; DUT, device under
test; SSPD, superconducting single-photon detector; TCSPC, time-correlated
single photon counting unit. RF Amp, RF amplifier; SG, signal generator;
PS, phase shifter.}

\label{FigS3}
\end{figure}

The single-photon characterization of the frequency shifting is realized
with the setup shown in Fig.$\,$\ref{FigS3} \cite{key-S2}. Strong
1550$\,$nm optical pulses from the mode-lock laser are sent into
a periodically-poled lithium niobate (ppLN) crystal waveguide to generate
775$\,$nm optical pulses using second harmonic generation (pulse
duration \textasciitilde{} 2$\,$ps). Then the 775$\,$nm optical
pulses are coupled into a periodically-poled potassium titanyl phosphate
(ppKTP) crystal waveguide with 10.5$\,$mm length and 110$\,\mu$m,
and 1550$\,$nm photon pairs are generated through type-II spontaneous
parametric down conversion. The spectral correlation between photon
pairs can be neglected \cite{key-S3}. As the generated photon pairs
have orthogonal polarization, we use polarization beam splitter (PBS)
to determinstically separate photons of each pair. Optical bandpass
filters (OBPF) are used in both optical paths to ensure only photons
with longer wavelength are coupled into the device, and only photons
with shorter wavelength are coupled into the optical delay line (ODL).
In the end, photons are coupled into a 50:50 beam splitter, and detected
by two superconducting single-photon detectors (SSPD) in a cryostat
cooled to 1.65$\,$K \cite{key-S4}. The photon incoming events of
the two SSPDs are recorded by the time-correlated single photon counting
unit (TCSPC). By comparing the arrival time list of the two SSPDs,
we can identify the coincidence events from the correlated photon
pairs. The coincidence rate is calculated based on a 256$\,$ps binning
of the photon arrival time. In the whole process, the 1550$\,$nm
optical pulses from the mode-lock laser are synchronized with the
RF drive for the mechanical mode with the method shown in Fig.$\,$S1.
Therefore, the photon pairs generated in ppKTP are also synchronized
with the RF drive.

\begin{figure}[H]
\begin{centering}
\includegraphics[width=0.25\paperwidth]{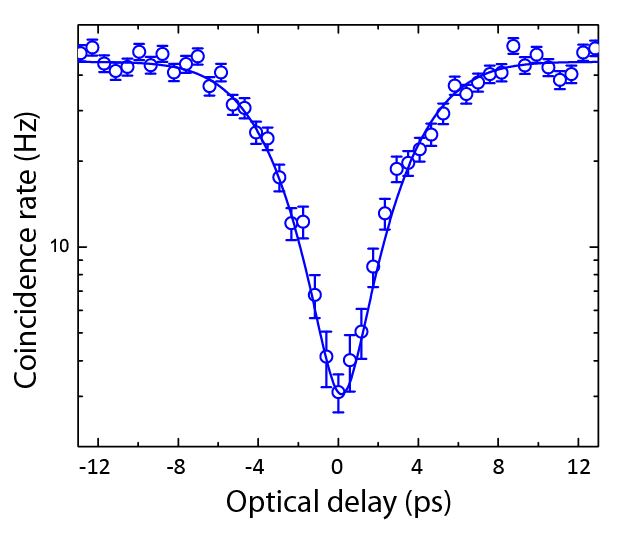}
\par\end{centering}

\caption{\textbf{Coincidence rate between the two SSPDs using photon pairs
with the same wavelength}}

\label{FigS4}
\end{figure}

We also performed the quantum interference experiment without the
device. By tuning the wavelength of optical pulses from the mode-lock
laser, photon pairs with the same wavelength can be generated. Using
photon pairs with the same wavelength, we observe the maximum $V=(92.4\pm4.3)\%$
visibility of the coincidence rate (Fig.$\,$\ref{FigS4}).

\section{Photon propagation loss in the frequency shifter}

The photon propagation loss in the frequency shifter consists of the
intrinsic material loss and the scattering loss induced by waveguide
pedestals. The intrinsic material loss of AlN waveguide is around
0.6$\,$dB/cm, contributing 0.36$\,$dB to the total propagation loss
of the 6$\,$mm long device \cite{key-S5}. The scattering loss is
less than 0.005$\,$dB per pedestal, which is determined through finite-element
simulations, thus the total scattering loss is less than 0.3$\,$dB.
As a result, the total propagation loss is less than 0.66$\,$dB.

\end{document}